\documentclass[%
 reprint,
superscriptaddress,
frontmatterverbose, 
nofootinbib,
nobibnotes,
 amsmath,amssymb,
 aps,
prb,
longbibliography]{revtex4-2}
\usepackage{graphicx}
\usepackage{dcolumn}
\usepackage{bm}
\usepackage{hyperref}
\hypersetup{colorlinks = true, citecolor = blue, linkcolor=blue, urlcolor=blue, pdfauthor=author}
\begin{document}
\preprint{APS/123-QED}

\title{Symmetrical Anisotropy Enables Dynamic Diffraction Control in Photonics}

\author{Hicham Mangach}
\affiliation{%
Light, Nanomaterials Nanotechnologies (L2n), CNRS-ERL 7004, Université de Technologie de Troyes,10000 Troyes, France\\
}%
\affiliation{%
Laboratory of optics, information processing, Mechanics, Energetics and Electronics, Department of Physics, Moulay Ismail University, B.P. 11201, Zitoune, Meknes, Morocco\\
}%
\author{Youssef El Badri}
\affiliation{%
Laboratory of optics, information processing, Mechanics, Energetics and Electronics, Department of Physics, Moulay Ismail University, B.P. 11201, Zitoune, Meknes, Morocco\\
}%

\author{Abdenbi Bouzid}
\affiliation{%
Laboratory of optics, information processing, Mechanics, Energetics and Electronics, Department of Physics, Moulay Ismail University, B.P. 11201, Zitoune, Meknes, Morocco\\
}%

\author{Younes Achaoui}
\affiliation{%
Laboratory of optics, information processing, Mechanics, Energetics and Electronics, Department of Physics, Moulay Ismail University, B.P. 11201, Zitoune, Meknes, Morocco\\
}%

\author{Shuwen Zeng }
\email{{Corresponding author: shuwen.zeng@cnrs.fr}}
\affiliation{%
Light, Nanomaterials Nanotechnologies (L2n), CNRS-ERL 7004, Université de Technologie de Troyes,10000 Troyes, France\\
}%

\date{\today}

\begin{abstract}
Despite the steady advancements in nanofabrication made over the past decade that had prompted a plethora of intriguing applications across various fields, achieving compatibility between miniaturized photonic devices and electronic dimensions remains unachievable due to the inherent diffraction limit of photonic devices. Several approaches have emerged to overcome the diffraction restriction and leverage the spatial information carried by the evanescent waves. Negative dielectric permittivity materials can be utilized to build photonic crystals (PhCs) based on surface plasmon-polaritons. This approach, however, is known to be exceedingly dissipative, leading to significant optical losses for photonic components. Herein, we report an approach based on the anisotropic scaling of the shapes of PhCs to impede the diffraction barrier and enable a tunable diffraction limit. This approach opens up avenues for high-frequency wave guiding in cermet configuration, which was previously unachievable. Furthermore, asymmetric and symmetric dimer network-type PhCs were explored, with the asymmetric case demonstrating a quasi-bound state in the continuum with a quality factor of up to 41000.
\end{abstract}

\maketitle


\section{\label{sec:level1}Introduction}

Mastering the art of wave manipulation has been a centuries-long endeavor, starting with the elderly Greeks to James Clack Maxwell, who sat forth the fundamental framework that underlies our modern understanding of the activity of electromagnetic waves in a medium \cite{watts2012metamaterial}. Over the last two decades, the advancements made in theory and experiments have enabled the feasibility of endowing artificial materials with wave-handling functionalities beyond the ultimate limitations found in nature, thereby setting a revolutionary milestone in the discipline of optics \cite{Shalaev:05, valentine2008three}. The emergence of quantum band theory of solids, which stipulates that electronic waves interact with periodically arranged quantum barriers to form a forbidden energy bands, was the earliest spark in the development of these structured materials \cite{john1987strong, sigalas1993band}. Photonic crystals (PhCs) were proposed and thoroughly investigated thereafter \cite{yablonovitch1987inhibited,krauss1996two}. The notion of one-dimensional stop bands, however, dates back to Lord Rayleigh’s demonstration in $1887$ that an infinitesimal periodic modulation of the material density within a structure may generate a narrow directional band gap, resulting in total reflection \cite{rayleigh1887xvii}. The meteoric rise of this concept was at the forefront of designing a slew of applications wherein the device functionalities are derived from the periodicity of the comprising units. Photonic waveguides, sensors, and graded-index lenses are some of the noteworthy achievements of such synthetic materials \cite{lonvcar2000design,estevez2012integrated,kurt2007graded}. Furthermore, miniaturizing dielectric photonic components to attain dimensions comparable to microelectronics has been significantly impacted by diffraction. This makes it extremely difficult to confine light into nanoscale regions smaller than the wavelength, thus significantly decreasing their effectiveness \cite{ozbay2006plasmonics,quan2019nanowires}.\\
According to the Helmholtz equation, the cavity length of the dielectric slab should be longer than $l_0$ \cite{doronin2022overcoming,born2013principles}, which makes subwavelength confinement of light infeasible in dielectric structures. 
Furthermore, Eq.~(\ref{eq:01}) states that the cavity length is inversely proportional to the refractive index, except that only materials with low refractive indices are commonly abundant in nature.
\begin{equation}\label{eq:01}
l_0=\frac{\lambda}{2\sqrt{\varepsilon_r}}
\end{equation}
Ebbesen et $al$.'s pioneering work revealed for the first time in $1998$ an extraordinary transmission of light through a perforated metal plate. Their findings showed an unexpected amplification of the transmitted wave beyond the diffraction limit through subwavelength apertures, representing a significant advancement in the field of optics \cite{ebbesen1998extraordinary}. Recently, negative dielectric permittivity has provided a new paradigm by exploiting subwavelength plasmonic resonance that not only precludes the diffraction barrier but also enhances the electromagnetic energy confinement at the nanoscale \cite{gramotnev2010plasmonics}. This salient feature has been used to resolve the information carried out through the spatial frequency of evanescent waves, therefore enabling photonic devices to be miniaturized beyond those conventionally available and attain sub-diffraction-limited resolution \cite{lu2012hyperlenses,barnes2003surface}. However, despite all the efforts made by plasmonic materials to bridge the gap between conventional PhCs and nanodevices, this approach still suffers from significant loss dissipation \cite{kuznetsov2016optically}. Nanophotonics based on surface plasmon polaritons (SPPs) offer an appropriate platform for achieving guided SPP modes beyond Abbe’s diffraction limit with subwavelength localization of energy, capable of operating both above and below the diffraction limit \cite{quan2019nanowires}. By the dawn of the twenty-first century, the emergence of the notion of left-handed metamaterials brought forth tremendous potential in perfect lenses, cloaking invisibility, perfect absorbers, and subwavelength resolution imaging \cite{pendry2000negative,schurig2006metamaterial,landy2008perfect,khorasaninejad2016metalenses}. Left-handed metamaterials make single and double negativities accessible through the periodic arrangement of locally resonant meta-atoms  \cite{veselago1967properties}. Nonetheless, the inherent complexity of the fabrication process, particularly for 3D nano-architectures, and the fact that this approach is only effective around the resonance frequency, remain some of the major hurdles facing these artificially engineered materials \cite{wegener2013metamaterials}. 
In this study, we report on a novel approach denoted as the anisotropic scaling effect (ASE) to elevate the diffraction limit in dielectric PhCs. We illustrate analytically the potential modifications imposed by the ASE on the photonic behavior expressed via a modified formalism of the eigenvalue problem. Moreover, PhCs are divided into two major categories from a topological standpoint: cermets that are made up of isolated high-density blocks embedded within a low-density host matrix, and networks, which are a density inversion of the previous category in which the patterns are connected to create a more compact structure \cite{economou1993classical}. In the subsequent sections, we elucidate the potential to elevate the diffraction barrier in these two configurations through the implementation of the ASE, which avoids the diffraction limitations along a one direction. A Finite Element Analysis (FEA) is conducted to assess the impact of ASE on the photonic dispersion curves and transmission spectra in both cermet and network configurations. Furthermore, PhCs composed of symmetric and asymmetric dimer of network-type are designed and optimized. The asymmetric dimer PhCs is found to exhibit a quasi-bound state in the continuum, with a high quality factor in its transmission response at the point of resonance. This funding highlights the potential suitability of the asymmetric dimer PhCs as a promising candidate for meeting the requirements of optical sensing applications.

\section{\label{sec:level2}Analytical Modeling}

To establish an analytical description of the scaling effect, we reiterate that anisotropically engineering the unit cells, i.e., introducing a unidirectional scaling factor $\alpha$ onto the geometry, equates to physically scaling the dielectric permittivity distribution along the scaling axis by the same factor. We considered scaling the \textit{y}-direction and leaving the other directions unaltered. Since the dielectric permittivity is scaled by the same factor as the geometry, a proportionate anisotropic scaling effect is thus introduced. After performing a variable modification of $r^{'}=r\alpha$, with $r^{'}=\sqrt{x^{2}+y^{'2}+z^{2}}$, the resulting dielectric permittivity and differential operator are written as follows:
\begin{equation}\label{eq:02}
\begin{aligned}
\varepsilon^{'}_{r}(r^{'})=\varepsilon_{r}(r^{}.\alpha)
\end{aligned}
\end{equation}
\begin{equation}\label{eq:03}
\begin{aligned}
\nabla^{'}=\frac{d}{d_{x}}+\frac{1}{\alpha}\frac{d}{d_{y^{'}}}+\frac{d}{d_{z}}
\\
\end{aligned}
\end{equation}
By inserting the new differential operator and dielectric function into the Helmholtz equation, a new eigenvalue problem for a nonmagnetic dielectric medium is described \cite{JOANNOPOULOS1997165}:
\begin{subequations}
\begin{equation}\label{eq:04a}
\frac{1}{\varepsilon^{'}_{r}(r^{'})}\alpha\nabla^{'}\times[\alpha\nabla^{'}\times E(r^{'},t)]=\frac{\omega^{2} }{c^{2}}E(r^{'},t)
\end{equation}
\begin{equation}\label{eq:04b}
\frac{1}{\varepsilon^{'}_{r}(r^{'})}\nabla^{'}\times[\nabla^{'}\times E(r^{'},t)]=(\frac{\omega }{\alpha c})^{2}E(r^{'},t)
\end{equation}
\end{subequations}
$\varepsilon_{r}(r^{'})$: Relative dielectric permittivity,\newline
$\omega, c$: Angular frequency and speed of light in vacuum, respectively.
The optical wave propagates along the {\textit{x}}-axis, which implies that it only experiences the unaltered modulated medium. Furthermore, the equation indicates that the new frequency $\omega^{'} =\frac{\omega }{\alpha}$ is greater since the scaling factor is less than unity. Thus, theoretically confirming that the frequency is scaled by ${\alpha}$, resulting in the shifting of the diffraction limit towards higher frequencies. The Floquet-Bloch theorem was then applied along the {\textit{x}} and {\textit{y}}-axes to investigate the 2D PhC model. According to the latter theorem, the solution of a periodic potential can be expressed as the product of a plane wave and a periodic function with the same periodicity as the crystal \cite{russell1986optics}.
\begin{equation}\label{eq:05}
\Psi(r^{'}) = E(r^{'})u(r^{'})
\end{equation}
$u(r^{'})= u(r^{'}+a)$: A periodic function with the same period as the crystal.\newline
Eq.~(\ref{eq:05}) presents the mathematical formulation of the Floquet-Bloch theorem, which may encounter challenges associated with negative eigenvalues. To overcome this concern, one can employ the Fourier basis and represent the periodic function in the form of a Fourier series, as illustrated by Eq.~(\ref{eq:06}):
\begin{equation}\label{eq:06}
f(r^{'})=\sum_{G} f_{G}exp(iGr^{'}) 
\end{equation}

Where, $G$ symbolizes a set of reciprocal lattice vectors. However, our study specifically focuses on transverse electric waves with an out-of-plane electric field component denoted by $E_z$.\\

\section{\label{sec:level3}Network Configuration}

\begin{figure}[!ht]
\includegraphics[width= 8.6 cm]{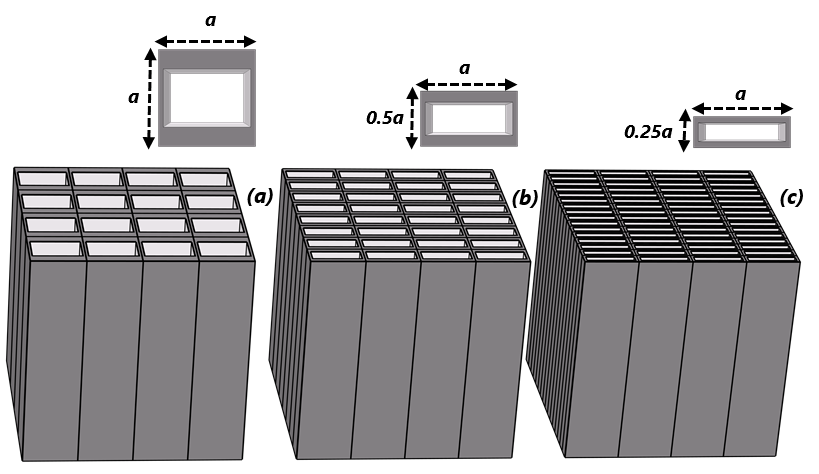}
\caption{\label{fig:01} Illustrations of the PhC structures studied: (a) Conventional silicon PhC with symmetric unit cell. (b) and (c) the halved (0.5\,a) and quartered (0.25\,a) PhCs, respectively.}
\end{figure}

For the sake of delineating the feasibility of utilizing the anisotropic scaling effect to produce network-type PhCs with an adjustable diffraction limit frequency, we first consider a typical 2D PhC with a spatial period {\textit{a}}, and an air-gap rectangular defect with a preset filling factor of {\textit{f}}, as indicated in Table.~\ref{tab:table1}. The unit cell is perfectly symmetrical along the {\textit{x}} and {\textit{y}}-axes, as depicted in [Fig.~\ref{fig:01}(a)], while the rectangular air-gap in this case has a width of {\textit{0.6\,a}} and a height of {\textit{0.9\,a}}.
\begin{table*}[ht!]
\caption{\label{tab:table1} Structural characteristics of network-type PhCs in normal, halved, quartered, symmetric, and asymmetric dimer configurations.}
\begin{ruledtabular}
\begin{tabular}{ccccccc}
 $a$ & $f$ & $scaling$ & $T_1$ & $T_2$ & $N_x$ & $N_y$
\\ \hline
 $1[\mu m]$ & $ 0.54$ & $1, 1/2, 1/4$ & $100[nm]$ & $30[nm]$ & $7$ & $1$ \\
\end{tabular}
\end{ruledtabular}
\end{table*}
 It will serve as a reference for all subsequent numerical simulations conducted in this section. Concerning the network-type PhCs, we opted for silicon material as the high-optical density host matrix and air as low-optical density inclusions to form the network pattern. Furthermore, silicon is produce on a large-scale in the industry, and it is considered one of the most essential materials on the frontiers of modern technology. Owing to its potential in on-chip photonic device manufacturing for light confinement, silicon PhC technology has matured to the point where it may outperform the photovoltaic and electrical industries \cite{priolo2014silicon,hochberg2010towards}. For all the reasons stated above, we have taken a keen interest in building our structures with monocrystalline silicon. All materials used in this work are assumed to be dispersive and loss free. The refractive index of silicon in the relevant region is approximately $n_{Si} = 3.5$, whereas that of air is assumed to be a constant value of $n_{Air} = 1$.\\ 

\begin{figure}[!ht]
\includegraphics[width= 8.6 cm,height=6 cm]{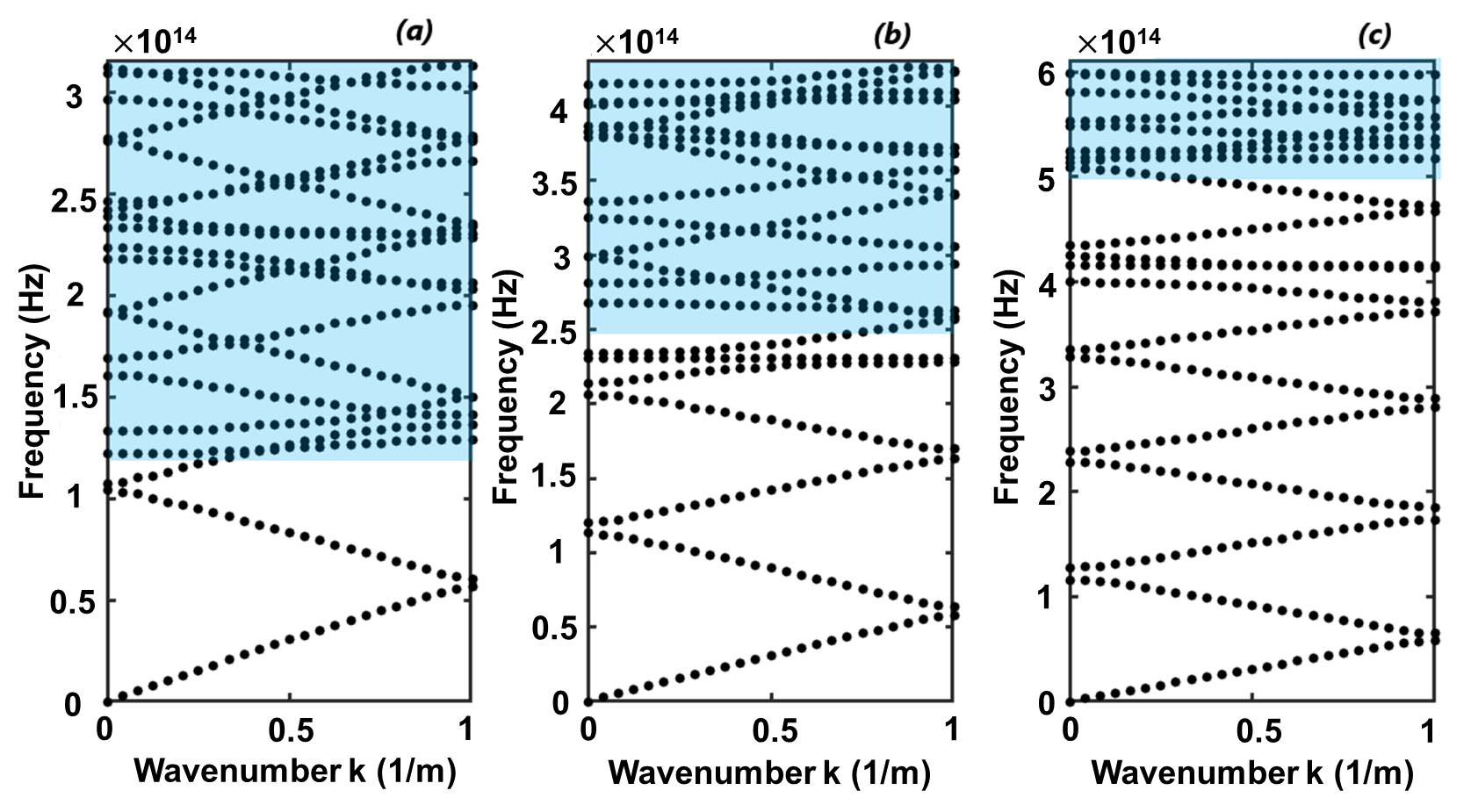}
\caption{\label{fig:02} Dispersion diagrams of conventional (a), halved (b) and quartered (c) network-type PhCs along the \( \Gamma X \) direction with a wavenumber of $k_{x} \in [0,1 \times \pi/ a]$.}
\end{figure}

To evaluate the transmission spectra, we build an array of $N_x \times N_y$ unit cells with an electromagnetic wave propagating freely in the two regions on the left and right sides of the array. We applied a harmonic excitation source $E_z$ at the left side of the {\textit{x}}-axis and a detector on the right side. A periodic condition was also applied in the {\textit{y}}-direction to assume that the crystal is infinite along the perpendicular direction of the wave's propagation. Bragg diffraction is a restriction that greatly affects the performance of engineered periodic structures, effectively preventing their operation in the high-frequency range. [Fig.~\ref{fig:02}(a)] serves as an illustrative example, displaying the appearance of the diffraction curtain approximately after the first band folding. Consequently, conventional photonic devices can only operate at low frequencies, located below the diffraction barrier. Extending their operation range to much higher frequencies remains a priority, and this attribute is accomplished only through the development of subwavelength structures, which is a result of technological progress. Indeed, several strategies have been developed to sidestep this barrier, harnessing both the available materials in nature and the range of artificially manufactured ones. Surface plasmon-based photonics, which combine photonic characteristics with electronics miniaturization \cite{ozbay2006plasmonics}, and photonic nanojets with waists smaller than the diffraction limit, which allow light to pass through without significant diffraction \cite{chen2004photonic}, have been reported. [Fig.~\ref{fig:02}(b)] and [Fig.~\ref{fig:02}(c)] illustrate the diffraction curtain being elevated by a factor of two and four in the cases of the halved and quartered network-type PhCs, respectively.\\

\begin{figure}[!ht]
\includegraphics[width= 8.6 cm,height=6 cm]{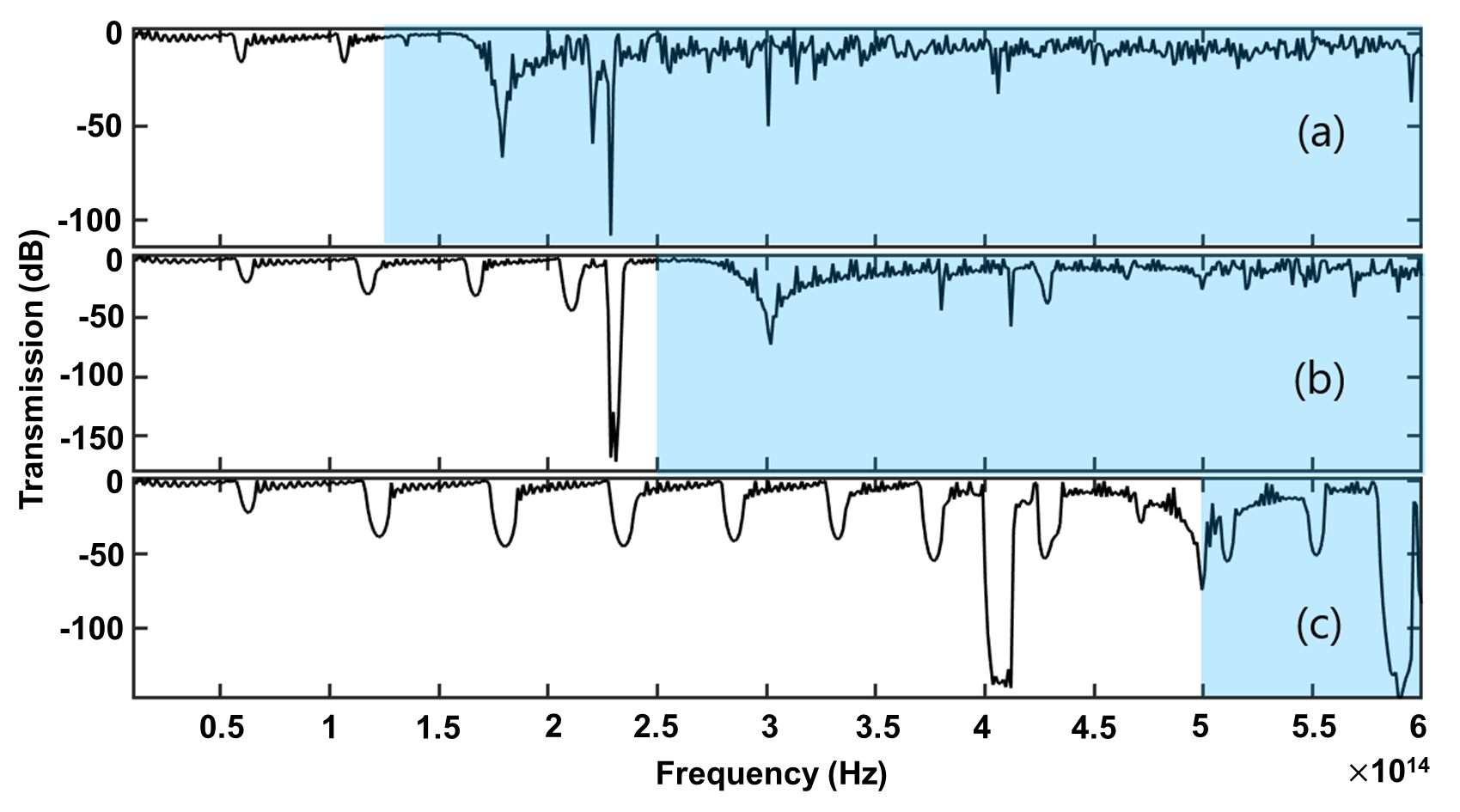}
\caption{\label{fig:03} Transmissions of conventional PhC at the top, halved PhC in the middle, and quartered PhC at the bottom of the network-type along the \( \Gamma X \) direction.}
\end{figure}
The corresponding transmission spectra of these network-type PhCs are depicted in [Fig.~\ref{fig:03}]. These results are in good accordance with the dispersion diagrams, where the blue regions indicate the Bragg limit position for each of the introduced structures. As a consequence, the diffraction limit in network-type PhCs occurs at $1.25\times 10^{14}$ Hz for the normal structure, $2.5\times 10^{14}$ Hz for the halved structure, and $5\times 10^{14}$ Hz for the quartered structure.
\begin{figure}[!ht]
\includegraphics[width= 8.6 cm]{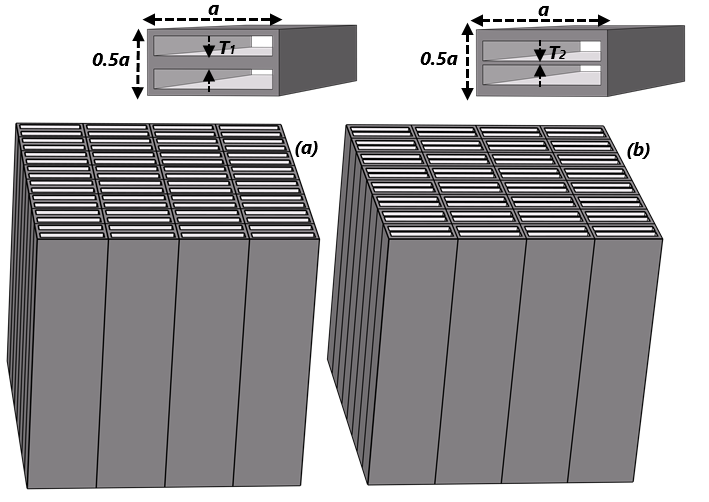}
\caption{\label{fig:04} Representations of the dimer network-type PhCs studied: (a) Symmetric dimer, and (b) Asymmetric dimer.
}
\end{figure}
Thus, our approach based on architectural engineering to surmount the diffraction constraint, although being unidirectional, provides a powerful tool for elevating the diffraction limit at will. [Fig.~\ref{fig:04}(a)] illustrates the design of an array of symmetric dimers. Each unit cell has a total width of half a period, obtained by arranging two quartered PhCs in the {\textit{y}}-direction. 
\begin{figure}[!ht]
\includegraphics[width= 8.6 cm,height=6 cm]{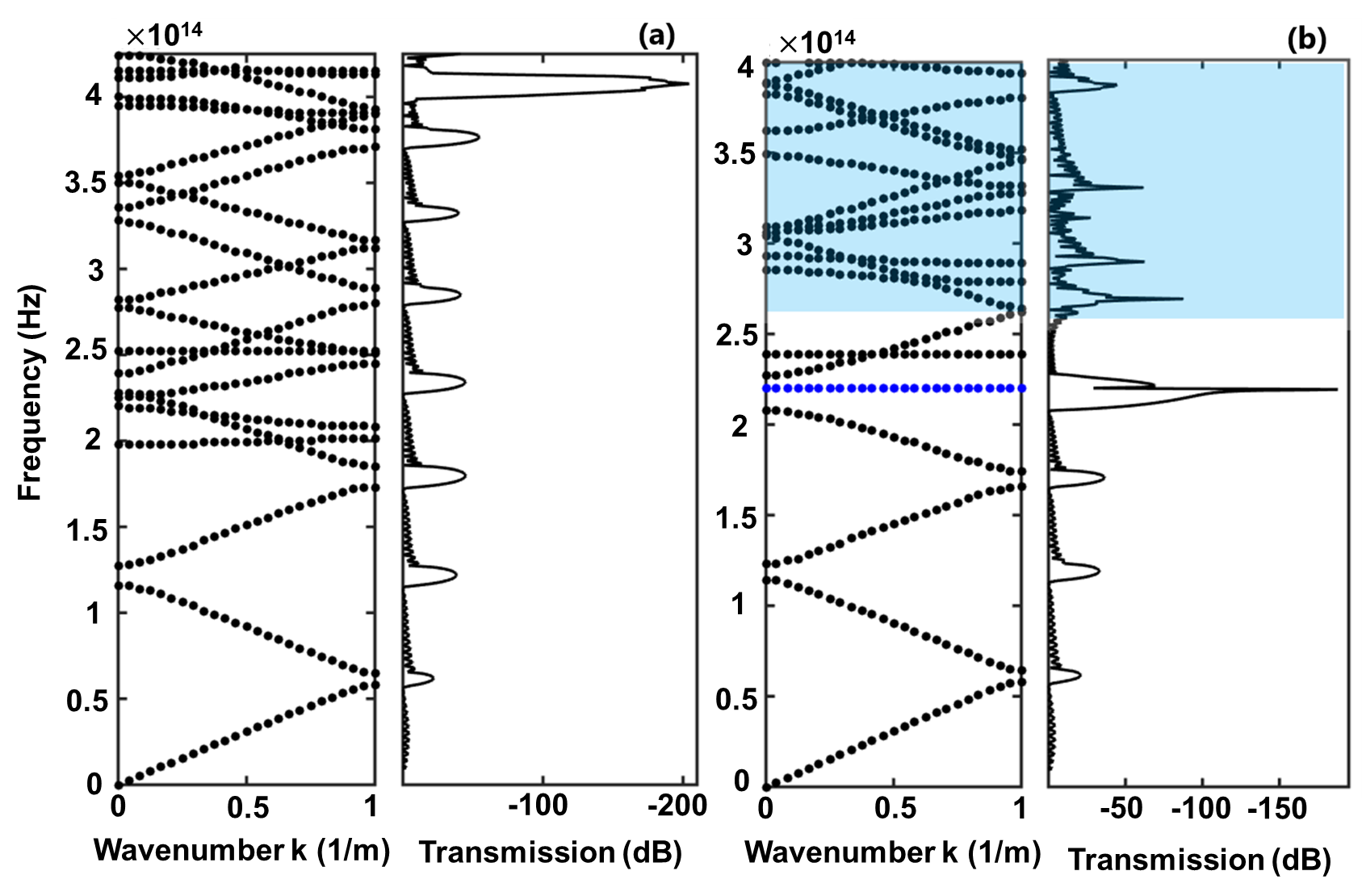}
\caption{\label{fig:05} Dispersion diagrams and associated transmission signatures of symmetric (a), and asymmetric (b) dimer network-type PhCs along the \( \Gamma X \) direction with a wavenumber of $k_{x} \in [0,1 \times \pi/ a]$.}
\end{figure}
We construct an array of asymmetric dimers by appropriately matching the spacing between the two air gaps within the symmetric dimers, as seen in [Fig.~\ref{fig:04}(b)]. The eigenvalue computations of symmetric and asymmetric dimers of network-type PhCs delineate that the symmetric case retains the primary feature of the quartered network-type PhC scenario, except that the optical wave encounters a unit cell with a half period, which results in a band overlap after the third band folding, as indicated in [Fig.~\ref{fig:05}(a)].
The transmission spectrum remains unaltered, confirming the hypothesis that band overlapping is unrelated to the diffraction phenomenon. 
\begin{figure}[!ht]
\includegraphics[width= 8.6 cm,height=7 cm]{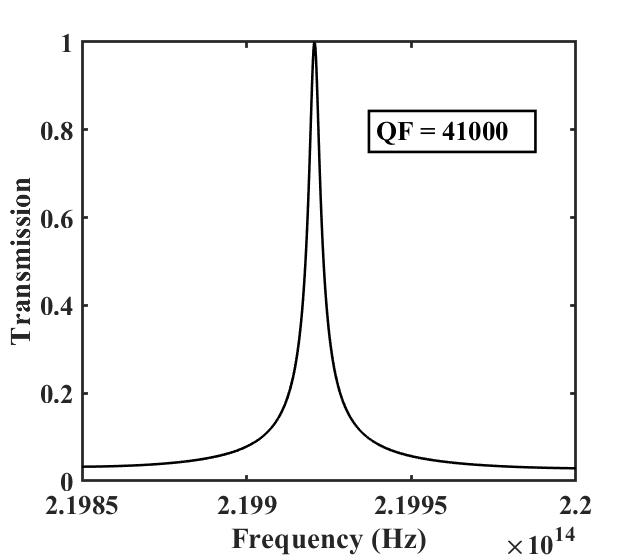}
\caption{\label{fig:06}Quality factor of the asymmetric dimer network-type PhC at the quasi-bound state.}
\end{figure}
The asymmetric dimer case exhibits a sharp transmission response due to the asymmetry of inclusions within it, which corresponds to a steady mode with a near-zero group velocity marked in blue, as can be seen in [Fig.~\ref{fig:05}(b)]. 
A more precise examination based on harmonic study to excite this localized mode is used to evaluate the associated quality factor, as shown in [Fig.~\ref{fig:06}]. This analysis uncovered a high quality factor of up to 41000 at normal incidence, which may seamlessly meet the stringent sensing requirements. 
Achieving such a high quality factor with a tiny modal volume equivalent to those attainable by photonic nanocavities based on energy confinement at subwavelength regime simply by implementing an asymmetric dimer is certainly interesting \cite{akahane2003high,song2005ultra}.\\

\section{\label{sec:level4}Cermet configuration}

\begin{figure}[!ht]
\includegraphics[width= 8.6 cm]{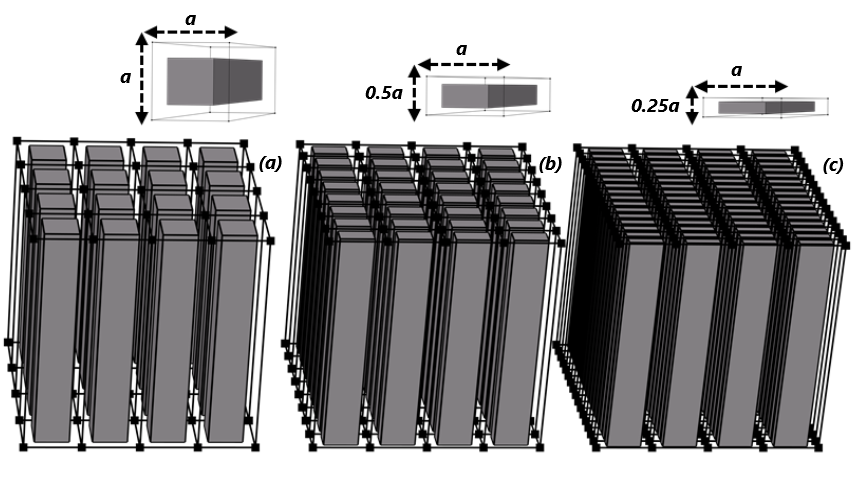}
\caption{\label{fig:07}Cermet-type configurations studied: Panel (a) represents the ordinary PhC, while panels (b) and (c) exhibit the halved and quartered PhCs, respectively.}
\end{figure}

In contrast to the network-type PhCs previously discussed, cermet-type PhCs have high-density optical inclusions (silicon) in a matrix with low-density optical material (air) \cite{lamb1980long}. Table.~\ref{tab:table2} outlines the new geometrical parameters of the cermet configuration. [Fig.~\ref{fig:07}(a)] depicts the cermet-type PhCs with the same spatial period as network-type PhCs. The width and height of silicon pillars are equal to {\textit{0.6\,a}}, whereas [Fig.~\ref{fig:07}(b)] and [Fig.~\ref{fig:07}(c)] illustrate the halved and quartered cermet configurations, respectively. Photonic band diagrams are constructed using a plane wave expansion to investigate the optical properties of such regularly spaced inclusions, and a harmonic analysis is carried out to assess the transmission spectra \cite{meade1993accurate}.
\begin{table}[h!]
\caption{\label{tab:table2} Geometrical parameters of Cermet-type PhCs configuration.}
\begin{ruledtabular}
\begin{tabular}{ccccc}
$a$ & $f^{'}$ & $scaling$ & $N'_x$ & $N'_y$ \\
\colrule
$1[\mu m]$ & $0.36$ & $1,1/2,1/4$ & $8$ & $11$ \\
\end{tabular}
\end{ruledtabular}
\end{table}

\begin{figure}[!ht]
\includegraphics[width= 8.6 cm,height= 5.5 cm]{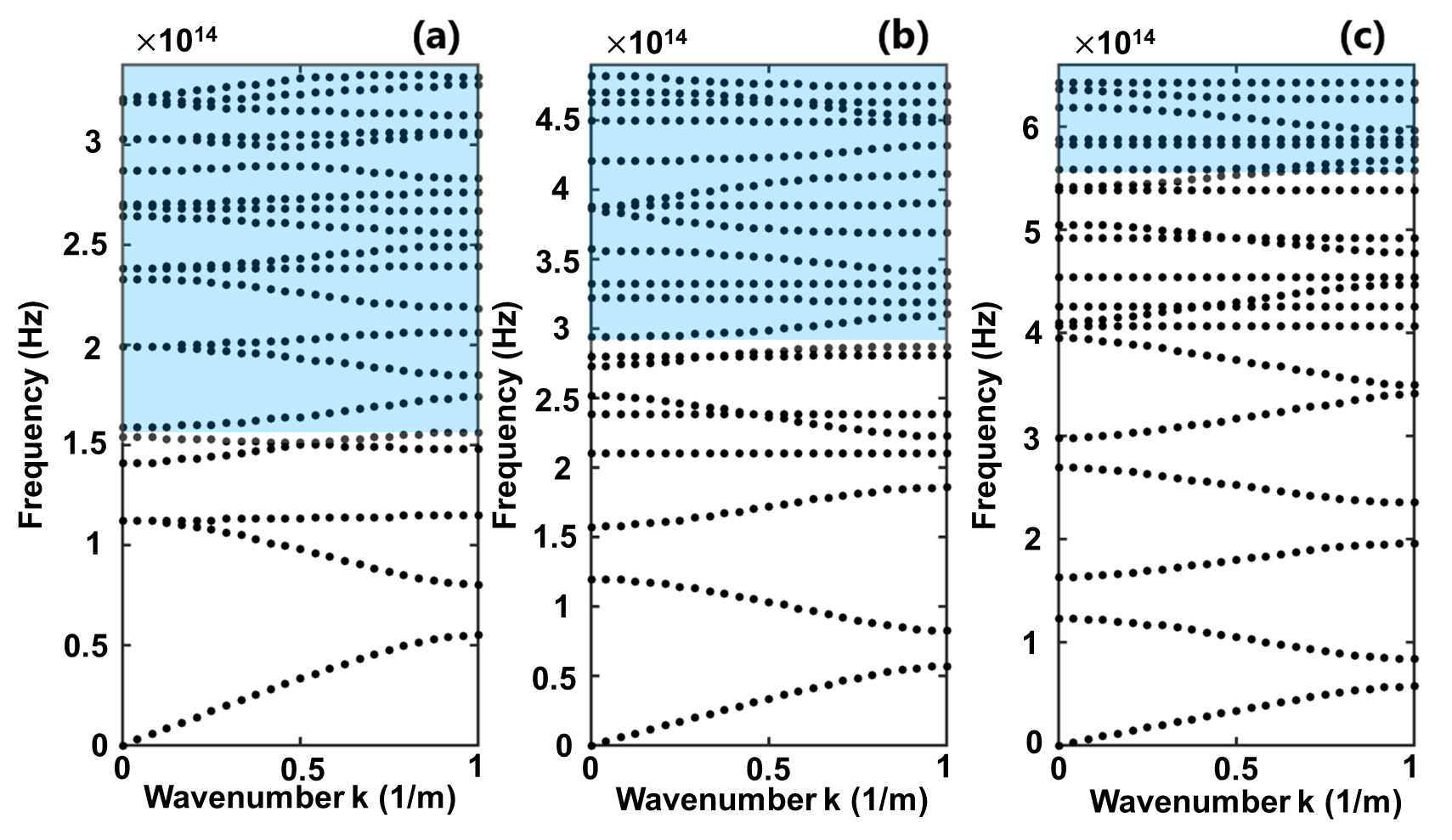}
\caption{\label{fig:08}The dispersion spectra of the cermet-type configuration: Panel (a) displays the ordinary PhC structure, while panels (b) and (c) depict the halved and quartered PhCs along the \( \Gamma X \) direction with a wavenumber of $k_{x} \in [0,1 \times \pi/ a]$, respectively.}
\end{figure}

\begin{figure}[!ht]
\includegraphics[width= 8.6 cm,height= 6.5 cm]{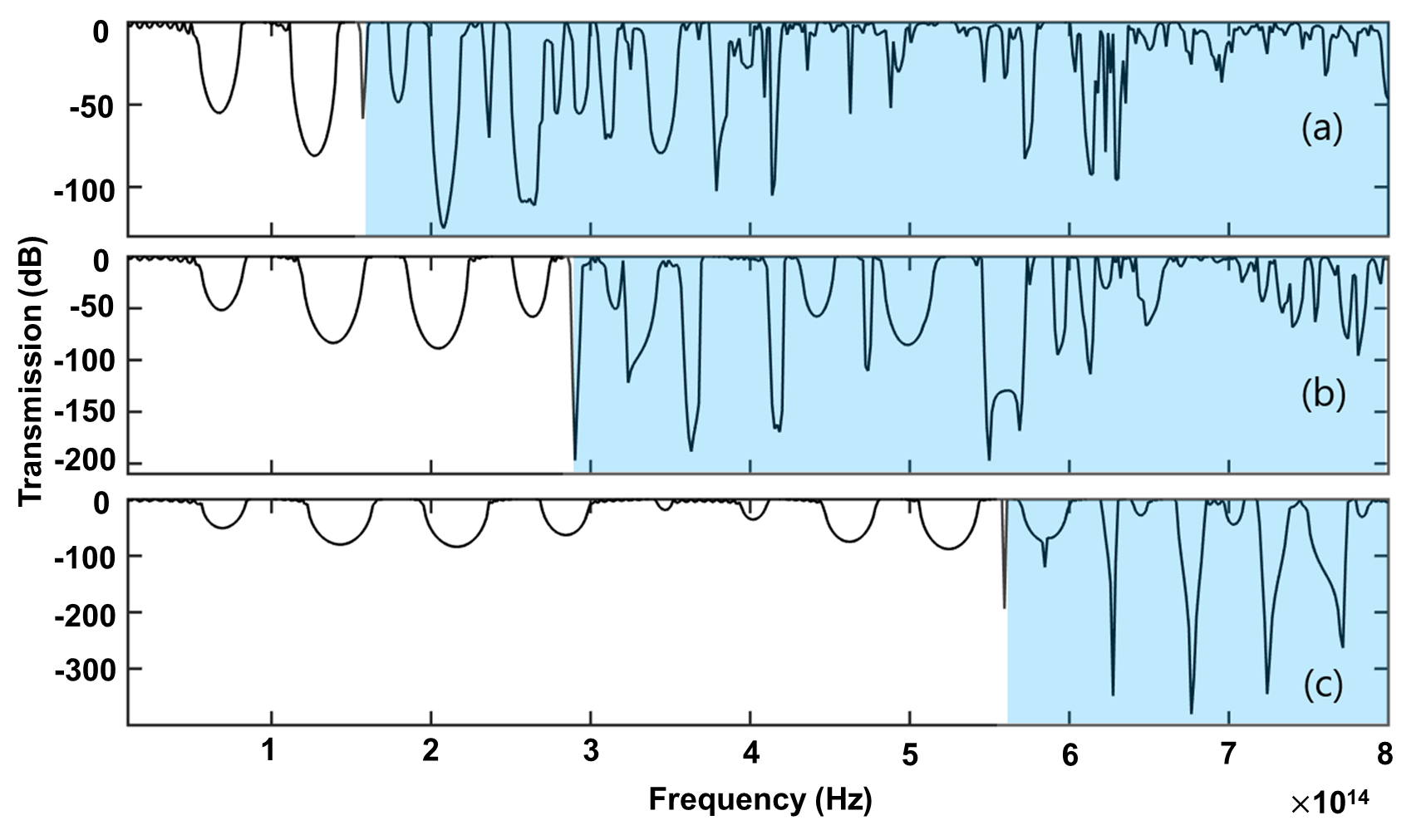}
\caption{\label{fig:09}The transmission signatures of the cermet-type configurations are presented as follows: (a) depicts the ordinary PhC structure, while (b) and (c) illustrate the halved and quartered PhC structures along the \( \Gamma X \) direction, respectively.}
\end{figure}

[Fig.~\ref{fig:08}] reveals the potential of the ASE to raise the diffraction curtain toward higher frequencies, while sustaining a large bandgaps within each band folding. The locations of the sharp decreases in transmission perfectly match the bandgaps for each trial, as shown in [Fig.~\ref{fig:09}]. Therefore, the transmission spectra are consistent with the dispersion diagrams, indicating that the Bragg limit in cermet-type PhCs jumps from $1.6\times10^{14}$ Hz, $ 2.9\times10^{14}$ Hz to $5.7\times10^{14}$ Hz for the standard, halved, and quartered cases, respectively.\\


\begin{figure}[!ht]
\includegraphics[width= 8.6 cm,height= 5.5 cm]{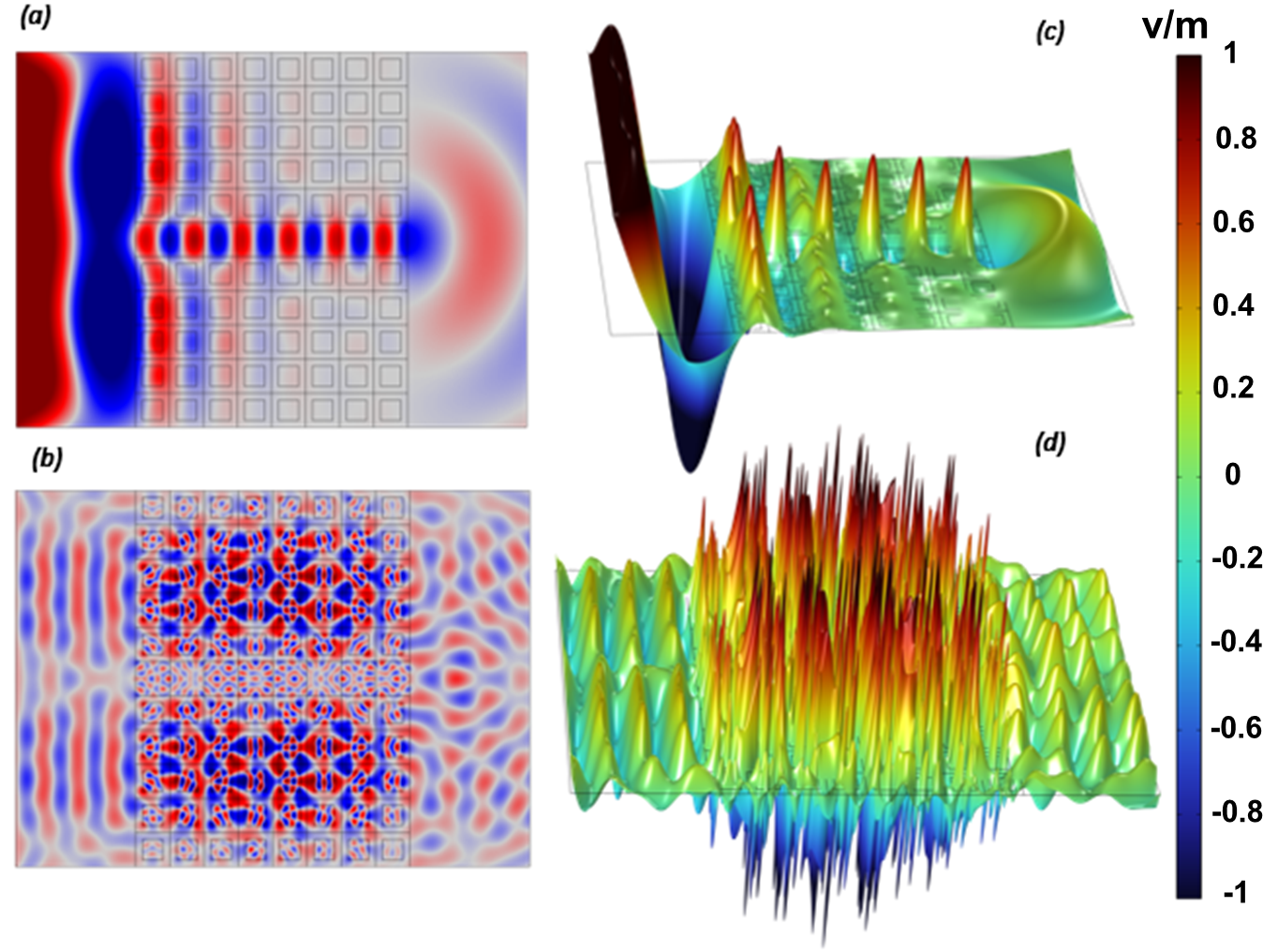}
\caption{\label{fig:10}The distributions of the out-of-plane electric field component and their high expressions are depicted for ordinary cermet-type PhC. Panels (a) and (c) correspond to the first band folding at a frequency of $0.7$ THz, while panels (b) and (d) illustrate the diffraction zone at a frequency of $3$ THz.}
\end{figure}

The dispersion diagrams of the cermet-type PhCs reveal large band gaps, which could have important implications for various applications, such as energy harvesting and electromagnetic wave guiding \cite{fan2001waveguide,soukoulis2013photonic}. Previous research on cermet-type PhCs has employed energy localization in the flat mode; however, our objective was to investigate the feasibility of using ASE to create cermet-type PhCs capable of effectively steering high-frequency electromagnetic waves, which were previously limited by the diffraction constraint. To this end, we constructed a superlattice consisting of $(N'_{x} \times N'_{y})$ unit cells of conventional, halved, and quartered cermet-type PhCs with defects along the {\textit{y}}-direction. [Fig.~\ref{fig:10}(a)] and [Fig.~\ref{fig:10}(b)] illustrate the behavior of guided and diffracted waves through the superlattice assembly, while [Fig.~\ref{fig:10}(c)]  and [Fig.~\ref{fig:10}(d)] show the corresponding out-of-plane electric fields at two distinct frequencies: $0.7$ THz and $3$ THz.\\
\begin{figure}[!ht]
\includegraphics[width= 8.6 cm,height= 5.5 cm]{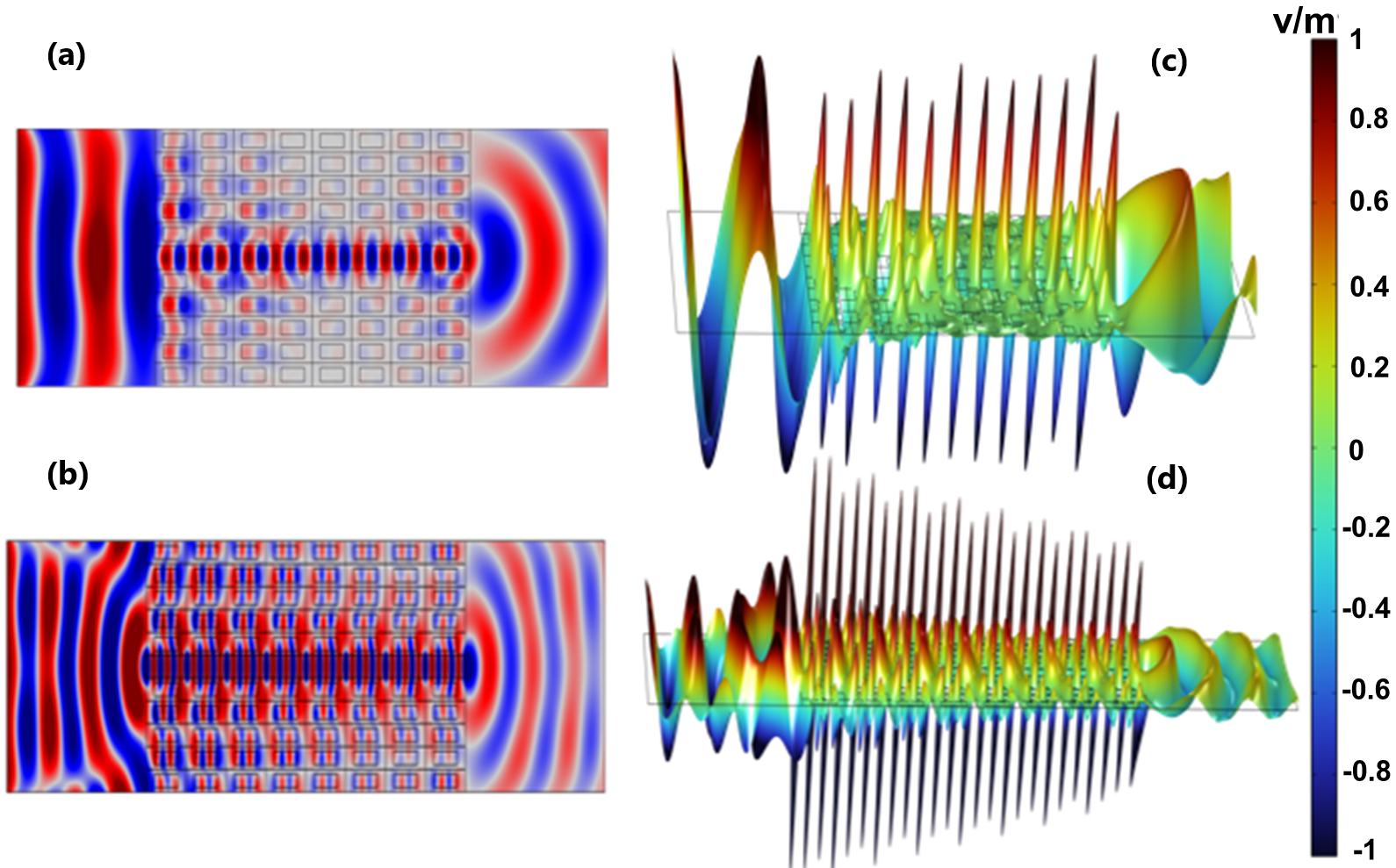}
\caption{\label{fig:11}The distributions and high expressions of the out-of-plane electric field component for halved and quartered cermet-type PhCs. Maps (a) and (c) depict the halved PhC at the second band folding frequency of $1.4$ THz, while maps (b) and (d) illustrate the quartered PhC at a frequency of $2.9$ THz.}
\end{figure}

The out-of-plane electric field $E_z$ is properly steered below the Bragg limit, while the plane wave is disrupted above this critical frequency, which captures the inadequacy of typical PhCs in conducting high-frequency photons. High-frequency guided waves are demonstrated employing our new suggested technique that avoids the diffraction limit. [Fig.~\ref{fig:11}(a)] and [Fig.~\ref{fig:11}(b)] demonstrate the capacity of the anisotropic effect to effectively guide optical waves in halved and quartered cermet-type PhCs at much higher frequencies, specifically at $1.4$ THz and $2.9$ THz, respectively. This allows overcoming the previously unavoidable diffraction constraint. The behavior of the out-of-plane electric field $E_z$ within the defect is provided for both cases in [Fig.~\ref{fig:11}(c)] and [Fig.~\ref{fig:11}(d)], indicating that the guided waves remain unaltered due to the introduction of the anisotropic geometry effect.\\

\section{\label{sec:level5}Conclusion}

This work reports on the use of an anisotropic architecting approach as a reliable way to circumvent the diffraction limit in two regular dielectric PhC configurations. This method has numerous far-reaching consequences and provides a new avenue for manipulating electromagnetic waves at high frequencies. It is also worth emphasizing that these findings can be expended to other disciplines, including phononics or elastodynamics, to attain tiny devices operating at much higher frequencies, thereby pushing beyond the levels of miniaturization currently available. Besides, both symmetric and asymmetric quartered dimer network-type photonic crystals were explored. The symmetric case remained diffraction-omitted, while the asymmetric case displayed a quasi-bound state in the continuum, resulting in a sharp peak in the transmission spectrum with a highly quality factor of up to 41000. This latter feature is congruent with optical sensing requirements, making it a promising candidate for accurate sensing applications.

\begin{acknowledgments}
We thank all of the authors for their contributions and helpful discussions. This work was supported by the UTT Project Stratégique NanoSPR (OPE-2022-0293), the Graduate School (Ecole Universitaire de Recherche) “NANOPHOT” (ANR-18-EURE-0013), PHC PROCORE-Campus France/Hong Kong Joint Research Scheme (No. 44683Q) and AAP1-LABEX SigmaPIX 2021.
\end{acknowledgments}


\nocite{*}
\begin{thebibliography}{}
\bibitem{watts2012metamaterial} C. M. Watts, X. Liu, and W. J. Padilla, Metamaterial
electromagnetic wave absorbers, \href{https://onlinelibrary.wiley.com/doi/abs/10.1002/adma.201200674}{Advanced Materials \textbf{24},OP98 (2012)}.
\bibitem{Shalaev:05} V. M. Shalaev, W. Cai, U. K. Chettiar, H.-K. Yuan, A. K. Sarychev, V. P. Drachev, and A. V. Kildishev, Negative
index of refraction in optical metamaterials, \href{ https://opg.optica.org/ol/abstract.cfm?URI=ol-30-24-3356}{Opt. Lett. \textbf{30},3356 (2005)}.
\bibitem{valentine2008three} J. Valentine, S. Zhang, T. Zentgraf, E. Ulin-Avila, D. A.
Genov, G. Bartal, and X. Zhang, Three-dimensional optical metamaterial with a negative refractive index, \href{https://doi.org/10.1038/nature07247 }{Nature \textbf{455},376 (2008)}.
\bibitem{john1987strong} S. John, Strong localization of photons in certain disordered dielectric superlattices, \href{https://link.aps.org/doi/10.1103/PhysRevLett.58.2486 }{Phys. Rev. Lett. \textbf{58},2486 (1987)}.
\bibitem{sigalas1993band} M. Sigalas and E. Economou, Band structure of elastic waves in two dimensional systems, \href{https://www.sciencedirect.com/science/article/pii/003810989390888T }{Solid State Communications \textbf{86},141 (1993)}.
\bibitem{yablonovitch1987inhibited} E. Yablonovitch, Inhibited Spontaneous Emission in Solid-State Physics and Electronics, \href{https://link.aps.org/doi/10.1103/PhysRevLett.58.2059 }{Phys. Rev. Lett. \textbf{58},2059 (1987)}.
\bibitem{krauss1996two} T. F. Krauss, R. M. D. L. Rue, and S. Brand, Two-dimensional photonic-bandgap structures operating at near-infrared wavelengths, \href{https://doi.org/10.1038/383699a0 }{Nature \textbf{383},699 (1996)}.
\bibitem{rayleigh1887xvii} L. Rayleigh, On the maintenance of vibrations by forces of double frequency, and on the propagation of waves through a medium endowed with a periodic structure, \href{ https://doi.org/10.1080/14786448708628074}{The London, Edinburgh, and Dublin Philosophical Magazine and Journal of Science \textbf{24},145 (1887)}.
\bibitem{lonvcar2000design} M. Lon\v{c}ar, T. Doll, J. Vu\v{c}kovi\'{c}, and A. Scherer, Design and Fabrication of Silicon Photonic Crystal Optical Waveguides, \href{https://opg.optica.org/jlt/abstract.cfm?URI=jlt-18-10-1402 }{J. Lightwave Technol. \textbf{18},1402 (2000)}.
\bibitem{estevez2012integrated} M. Estevez, M. Alvarez, and L. Lechuga, Integrated optical devices for lab-on-a-chip biosensing applications, \href{https://onlinelibrary.wiley.com/doi/abs/10.1002/lpor.201100025 }{Laser \& Photonics Reviews \textbf{6},463 (2012)}.
\bibitem{kurt2007graded} H. Kurt and D. S. Citrin, Graded index photonic crystals, \href{https://opg.optica.org/oe/abstract.cfm?URI=oe-15-3-1240 }{Opt. Express \textbf{15},1240 (2007)}.
\bibitem{ozbay2006plasmonics} E. Ozbay, Plasmonics: Merging Photonics and Electronics at Nanoscale Dimensions, \href{ https://www.science.org/doi/abs/10.1126/science.1114849}{Science \textbf{311},189 (2006)}.
\bibitem{quan2019nanowires} L. N. Quan, J. Kang, C.-Z. Ning, and P. Yang, Nanowires for Photonics, \href{ https://doi.org/10.1021/acs.chemrev.9b00240}{Chemical Reviews \textbf{119},9153 (2019)}.
\bibitem{doronin2022overcoming} I. V. Doronin, E. S. Andrianov, and A. A. Zyablovsky, Overcoming the Diffraction Limit on the Size of Dielectric Resonators Using an Amplifying Medium, \href{ https://link.aps.org/doi/10.1103/PhysRevLett.129.133901}{Phys. Rev. Lett. \textbf{129},133901 (2022)}.
\bibitem{born2013principles} M. Born and E. Wolf, Principles of optics: electromagnetic theory of propagation, interference and diffraction of light, (Elsevier, 2013).
\bibitem{ebbesen1998extraordinary} T. W. Ebbesen, H. J. Lezec, H. Ghaemi, T. Thio, and P. A. Wolff, Extraordinary optical transmission through sub-wavelength hole arrays, \href{https://doi.org/10.1038/35570 }{Nature \textbf{391},667 (1998)}.
\bibitem{gramotnev2010plasmonics} D. K. Gramotnev and S. I. Bozhevolnyi, Plasmonics beyond the diffraction limit, \href{ https://doi.org/10.1038/nphoton.2009.282}{Nature Photonics \textbf{4},83 (2010)}.
\bibitem{lu2012hyperlenses} D. Lu and Z. Liu, Hyperlenses and metalenses for far-field super-resolution imaging, \href{https://doi.org/10.1038/ncomms2176 }{Nature Communications \textbf{3},1205 (2012)}.
\bibitem{barnes2003surface} W. L. Barnes, A. Dereux, and T. W. Ebbesen, Surface plasmon subwavelength optics, \href{https://doi.org/10.1038/nature01937 }{Nature \textbf{424},824 (2003)}.
\bibitem{kuznetsov2016optically} A. I. Kuznetsov, A. E. Miroshnichenko, M. L. Brongersma, Y. S. Kivshar, and B. Luk’yanchuk, Optically resonant dielectric nanostructures, \href{https://www.science.org/doi/abs/10.1126/science.aag2472 }{Science \textbf{354},aag2472 (2016)}.
\bibitem{pendry2000negative} J. B. Pendry, Negative Refraction Makes a Perfect Lens, \href{ https://link.aps.org/doi/10.1103/PhysRevLett.85.3966}{Phys. Rev. Lett. \textbf{85},3966 (2000)}.
\bibitem{schurig2006metamaterial} D. Schurig, J. J. Mock, B. J. Justice, S. A. Cummer, J. B. Pendry, A. F. Starr, and D. R. Smith, Metamaterial Electromagnetic Cloak at Microwave Frequencies, \href{https://www.science.org/doi/abs/10.1126/science.1133628 }{Science \textbf{314},977 (2006)}.
\bibitem{landy2008perfect} N. I. Landy, S. Sajuyigbe, J. J. Mock, D. R. Smith, and W. J. Padilla, Perfect Metamaterial Absorber, \href{https://link.aps.org/doi/10.1103/PhysRevLett.100.207402}{Phys. Rev. Lett. \textbf{100},207402 (2008)}.
\bibitem{khorasaninejad2016metalenses} M. Khorasaninejad, W. T. Chen, R. C. Devlin, J. Oh,A. Y. Zhu, and F. Capasso, Metalenses at visible wavelengths: Diffraction-limited focusing and subwavelength resolution imaging, \href{https://www.science.org/doi/abs/10.1126/science.aaf6644 }{Science \textbf{352},1190 (2016)}.
\bibitem{veselago1967properties} V. G. Veselago, Properties of materials having simultaneously negative values of the dielectric and magnetic susceptibilities, \href{https://cir.nii.ac.jp/crid/1573387450661251840 }{Soviet Physics Solid State USSR \textbf{8},2854 (1967)}.
\bibitem{wegener2013metamaterials} M. Wegener, Metamaterials Beyond Optics, \href{https://www.science.org/doi/abs/10.1126/science.1246545 }{Science \textbf{342},939 (2013)}.
\bibitem{economou1993classical} E. N. Economou and M. M. Sigalas, Classical wave propagation in periodic structures: Cermet versus network topology, \href{ https://link.aps.org/doi/10.1103/PhysRevB.48.13434}{Phys. Rev. B \textbf{48},13434 (1993)}.
\bibitem{JOANNOPOULOS1997165} J. Joannopoulos, P. R. Villeneuve, and S. Fan, Photonic crystals, \href{https://www.sciencedirect.com/science/article/pii/S0038109896007168 }{Solid State Communications \textbf{102},165 (1997)}.
\bibitem{russell1986optics} P. S. J. Russell, Optics of Floquet-Bloch waves in dielectric gratings, \href{ https://doi.org/10.1007/BF00697490}{Applied Physics B \textbf{39},231 (1986)}.
\bibitem{priolo2014silicon} F. Priolo, T. Gregorkiewicz, M. Galli, and T. F. Krauss, Silicon nanostructures for photonics and photovoltaics, \href{https://doi.org/10.1038/nnano.2013.271 }{Nature Nanotechnology \textbf{9},19 (2014)}.
\bibitem{hochberg2010towards}  M. Hochberg and T. Baehr-Jones, Towards fabless silicon photonics, \href{https://doi.org/10.1038/nphoton.2010.172 }{Nature Photonics \textbf{4},492 (2010)}.
\bibitem{chen2004photonic}  Z. Chen, A. Taflove, and V. Backman, Photonic nanojet enhancement of backscattering of light by nanoparticles: a potential novel visible-light ultramicroscopy technique, \href{https://opg.optica.org/oe/abstract.cfm?URI=oe-12-7-1214 }{Opt. Express \textbf{12},1214 (2004)}.
\bibitem{akahane2003high} Y. Akahane, T. Asano, B.-S. Song, and S. Noda, High-Q photonic nanocavity in a two-dimensional photonic crystal, \href{https://doi.org/10.1038/nature02063 }{Nature \textbf{425},944 (2003)}.
\bibitem{song2005ultra} B.-S. Song, S. Noda, T. Asano, and Y. Akahane, Ultra-high-Q photonic double-heterostructure nanocavity, \href{https://doi.org/10.1038/nmat1320 }{Nature Materials \textbf{4},207 (2005)}.
\bibitem{lamb1980long} W. Lamb, D. M. Wood, and N. W. Ashcroft, Long-wavelength electromagnetic propagation in heterogeneous media, \href{https://link.aps.org/doi/10.1103/PhysRevB.21.2248 }{Phys. Rev. B \textbf{21},2248 (1980)}.
\bibitem{meade1993accurate} R. D. Meade, A. M. Rappe, K. D. Brommer, J. D. Joannopoulos, and O. L. Alerhand, Accurate theoretical analysis of photonic band-gap materials, \href{ https://link.aps.org/doi/10.1103/PhysRevB.48.8434}{Phys. Rev. B \textbf{48},8434 (1993)}.
\bibitem{fan2001waveguide} S. Fan, S. G. Johnson, J. D. Joannopoulos, C. Manolatou, and H. A. Haus, Waveguide branches in photonic crystals, \href{ https://opg.optica.org/josab/abstract.cfm?URI=josab-18-2-162}{J. Opt. Soc. Am. B \textbf{18},162 (2001)}.
\bibitem{soukoulis2013photonic} C. M. Soukoulis, Photonic band gaps and localization, \textbf{308}, (Springer Science \& Business Media, 2013).

\end{thebibliography}

\end{document}